\documentclass[prl,aps,amssymb,twocolumn,superscriptaddress,showpacs]{revtex4}
\usepackage{psfig}
\usepackage{graphicx}
\def\kp{{\bf k}_\perp}
\begin{document}

\title{Ballistic Conductance and Magnetism in Short Tip-Suspended Ni Nanowires} 
\author{Alexander Smogunov}
\affiliation{SISSA, Via Beirut 2/4, 34014 Trieste (Italy)}
\affiliation{
INFM, Democritos Unit\`a di Trieste, Via Beirut 2/4,~34014 Trieste (Italy)}
\affiliation{Voronezh State University, University Sq. 1, 394006 Voronezh (Russia)}
\author{Andrea Dal Corso}
\affiliation{SISSA, Via Beirut 2/4, 34014 Trieste (Italy)}
\affiliation{
INFM, Democritos Unit\`a di Trieste, Via Beirut 2/4,~34014 Trieste (Italy)}
\author{Erio Tosatti}
\affiliation{SISSA, Via Beirut 2/4, 34014 Trieste (Italy)}
\affiliation{
INFM, Democritos Unit\`a di Trieste, Via Beirut 2/4,~34014 Trieste (Italy)}
\affiliation{ICTP, Strada Costiera 11, 34014 Trieste(Italy)}

\date{\today}

\begin{abstract}
Electronic and transport properties of a short Ni nanowire suspended between two
semi-infinite ferromagnetic Ni leads are explored in the framework of density-functional theory.
The spin-dependent ballistic conductance of the nanowire is calculated using a scattering-based 
approach
and the Landauer-B\"uttiker formula. The total calculated conductance in units of $G_0 = 2e^2/h$ 
is around 1.6, in fairly good agreement with the broad peak observed around $\sim$ 1.5 for the
last conductance step in break junctions. Separating contributions from different spins, 
we find nearly 0.5 $G_0$ from the majority spin $s$-like channel, whereas the 
remaining minority spin conductance of 1.1 $G_0$ contains significant contributions 
from several $d$ states, but much less than 0.5 $G_0$ from $s$ states. 
The influence of the structural relaxation on the magnetic 
properties and the ballistic conductance of the nanowire is also studied.

\end{abstract}


\maketitle                                                                                             

\section{Introduction}

Modern experimental techniques involving ballistic conductance measurements 
in tip-based devices or in mechanically controllable break junctions allow 
the study of contacts of ultimate dimensions consisting of just a single atom, 
or of a short monatomic wire in the constriction. Investigations of metal 
nanocontacts during elongation show that the conductance decreases stepwise following 
atomic rearrangements. In noble metals (Cu, Ag, Au) and alkali metals (Li, Na, K) 
the last conductance step before breaking, most likely corresponding to a monatomic 
contact or a nanowire, takes a value close to $G_0$ ($G_0=2e^2/h$ is the 
conductance quantum)\cite{1} which corresponds to the free propagation 
of two $s$-like conduction electrons, one per spin. Transition 
metal nanocontacts (e.g., Ni, Co, Fe) whose atoms have a partially occupied $d$ shell, 
also display conductance steps but show less clear evidence for conductance quantization. 
Some conductance histograms for Ni \cite{oshima,ono}
and Fe \cite{ott} nanocontacts 
exhibit a well-defined first peak near 1 $G_0$; 
some others show a striking fractional peak at 
0.5 $G_0$ in Ni \cite{ono, rodrigues} and Co.\cite{rodrigues} Generally, however, low 
temperature break junction measurements on Ni, Co, and Fe nanocontacts 
find a broadly peaked conductance histogram well above 1 $G_0$. 
Untiedt {\it et al.}\cite{untiedt} reported a value at roughly 
2 $G_0$ (in Fe) and 1.3 $G_0$ (in Co and Ni) while Bakker {\it et al.}\cite{yanson} 
found a conductance peak at 1.6 $G_0$ in Ni.
 
A variety of methods have been developed to address theoretically the problem 
of electron transport in atomic scale conductors.\cite{lcao,palacios,wavelet,lang,rtm,
finitediff,lkkr,kkr,wortmann,calzolari,choi}. 
Choi and Ihm\cite{choi} in particular proposed a complex wave-vector band method 
particularly suitable for implementation within standard plane-wave ab-initio 
density functional pseudopotential approaches. Owing to their exponential decay, 
complex wave-vector Bloch states provide the spatially localized states crucially involved 
in the lead-nanocontact-lead wave-function matching. For application to transition 
metal nanocontacts, we recently presented\cite{ourprb} an extension of Choi and Ihm's method 
to ultrasoft pseudo-potentials, making the treatment of 
very localized $d$ states accessible with a plane-wave basis 
set. The method was applied to study the effect of a magnetization 
reversal on electron transport in infinite monatomic Co and Ni nanowires.\cite{ourprb} 
Interestingly, these calculations showed that a spin reversal can completely 
block all the $d$ channels at the Fermi energy, causing a large 
``ballistic magneto-resistance'' and leaving a conductance of only 
$\sim 1~G_0$ of the freely propagating $s$-like states of both spins. However,
the nanocontact geometry could not be addressed by such a tipless calculation,  
which omits completely the finite electron reflection at the 
contact. In this paper we use the same method to fully include the
tips, and explore the joint effects of the geometrical constriction
and of magnetism on the ballistic transport of a magnetic metal like Ni. 

For our purpose, we consider an idealized system consisting of a three-atom 
Ni monatomic nanowire placed between two semi-infinite Ni(001) bulk-like ``leads''. 
Actually, break junction experiments showed spontaneous formation of monatomic
nanowires only for the heavy $5d$ metal nanocontacts (Ir, Pt, and Au),\cite{chains} 
whereas only transmission electron microscopy showed evidence 
of such short nanowire contacts of lighter transition metals such as Co 
and Pd.\cite{rodrigues} Our calculation is intended as a case nanocontact 
study, rather meant to explore than to affirm one particular geometry and its effects.
We note that recently Bagrets {\it et al.}\cite{bagrets} performed similar 
calculations for the three-atom Co nanowire while Jacob {\it et al.}\cite{jacob} 
studied ballistic transport in a single-atom Ni contacts using a 
cluster embedded method;\cite{palacios} 
we shall later compare our results with their calculations.

\section{Calculations: Geometry, Relaxation, and Magnetism}

\begin{figure}
\includegraphics[width=8.5cm]{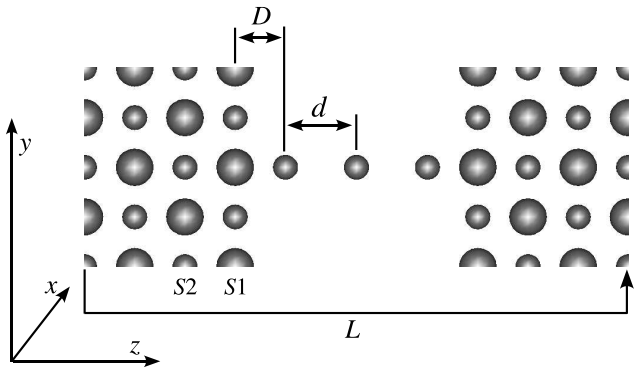}
\caption{\label{struc}
Supercell used for simulating a three-atom Ni wire suspended between two
semi-infinite Ni(001) leads.
}
\end{figure}

Electronic structure calculations are carried out within the spin density 
functional theory, using the standard plane-wave PWscf code.\cite{PWscf} 
The spin-polarized version of the generalized gradient 
approximation ($\sigma$-GGA) in the form introduced by Perdew, Burke, and 
Ernzerhof\cite{PBE} is used for exchange-correlation functionals.  
Ni atoms are described by ultrasoft
pseudo-potentials\cite{vanderbilt} with parameters given in
Ref.~\onlinecite{usparameters}. The cutoff kinetic energies were
25 Ry and 250 Ry for the wave functions and for the charge density, respectively.
Integration over the Brillouin Zone (BZ) up to the Fermi energy was performed by using 
special k points\cite{specialpoints} and a standard broadening technique\cite{smearing} 
with a smearing parameter of 0.02 Ry. 
Periodic boundary conditions are assumed in all three directions.
The supercell used to simulate a three-atom Ni nanowire contact is shown in Fig.~\ref{struc}.
A ($2\sqrt{2}\times2\sqrt{2}$) periodicity is employed in the $xy$ plane
which keeps the periodically repeated wires sufficiently 
apart from one another so that their mutual influence can be neglected.
The bulk ``tips'' are simulated by the two opposite (001) faces of a
planar slab consisting of seven atomic (001) crystalline 
planes (8 atoms per plane), quite sufficient to reproduce 
a bulk-like potential in the middle of the slab.
Since the supercell is very large along [001] (the $z$ axis), 
the BZ in this direction is sampled only at $k_z$ = 0 while in the ($k_x, k_y$) plane 
perpendicular to the wire convergence demanded instead ten two-dimensional (2D) 
special k points. This sampling of the BZ was found to be sufficient for obtaining 
a converged self-consistent potential needed for subsequent transmission calculations.

We started with an ideal unrelaxed atomic configuration
where the end atoms of the wire are positioned at the ideal hollow sites of
Ni(001) surfaces, their distance $D$ (see Fig. 1) to the surface planes being $a_0/2$ 
($a_0=3.52$~\AA~is the lattice constant of the bulk Ni).
The intra-wire spacing $d$ was initially set equal to $a_0/\sqrt2$
which is the nearest neighbor distance in the fcc Ni. 
We then relaxed structurally this configuration, using a simplified 
($2\times 2$) planar geometry (4 atoms per layer representing the bulk)
and fifteen special k points in the 2D BZ, which reduced the computational weight.
Minimizing total energy, all atomic positions were allowed to relax, 
including the distance parameters $D$, $d$, and the length $L$ of the supercell 
along the $z$ axis. During relaxation the wire was constrained
to remain straight and no zig-zag-like configurations were considered.
Some calculated quantities describing the geometry of the system are listed in Table~1.
In particular, we found that the first interlayer distance
in the leads is contracted with respect to the ideal unrelaxed value $d_{12}^0=a_0/2$
by value of $0.8-2.8\%$.\cite{nota1} These values are, however, less than the 
corresponding one 
(3.6\% $d_{12}^0$) calculated for the unperturbed, contact free Ni(001) surface\cite{mittendorfer} 
reflecting the influence of the nanowire bridging two surfaces.
As expected, the most significant geometry change occurs in the positions of the nanowire atoms.
Here, both the end distance $D$ and the central spacing $d$ 
contract significantly during the relaxation. The value of $d$ (2.19~\AA) is 
in fact very close to the calculated zero-stress interatomic spacing 
in the infinite monatomic wire (2.17~\AA).
The atomic positions obtained after relaxation were subsequently 
used to construct 
our final system with a larger ($2\sqrt{2}\times2\sqrt{2}$) planar periodicity,\cite{nota2} 
that would have been much more expensive to optimize directly.

\begin{table}
\caption{
Structure of the three-atom Ni wire suspended
between two Ni(001) leads in unrelaxed and relaxed configurations:
relaxation of the first interlayer spacing in the leads
($\Delta_{12}$), lead-nanowire separation ($D$), and interatomic distance
in the nanowire ($d$).
Some relevant theoretical data for the Ni(001)
surface and for the infinite Ni nanowire are also provided.}
\begin{ruledtabular}
\begin{tabular}{c|ccc}
{} & $\Delta_{12}$ (\%) & $D$ (\AA) & $d$ (\AA)
\\
\hline
unrelaxed wire & 0.0 & 1.76 & 2.49 \\
relaxed wire   & $-0.8\div-2.8$ & 1.47 & 2.19 \\
Ni(001) surface$^a$ & $-3.6$ & $-$ & $-$ \\
infinite Ni wire & $-$ & $-$ & 2.17 \\
\end{tabular}
\end{ruledtabular}
\label{table1}
\leftline{$^a$ Ref. \onlinecite{mittendorfer}}
\end{table}

\begin{figure}
\includegraphics[width=8.5cm]{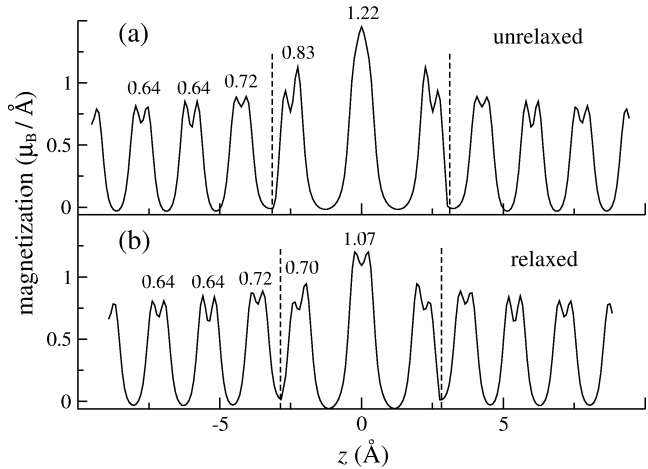}
\caption{\label{mag}
The $xy$ integrated magnetization as a function
of $z$ for the tip suspended three-atom Ni wire in unrelaxed (a) and relaxed (b) configurations.
Magnetic moments (in the Bohr magneton) at some atoms are also given.
}
\end{figure}

LSDA calculations were carried out for this geometry, comprising a total of 59 Ni atoms.
In Figs.~\ref{mag}a and \ref{mag}b we show the local magnetization after averaging 
in the $xy$ plane as a 
function of $z$ for unrelaxed and relaxed atomic configurations, respectively.
The total magnetization in the lead region (marked by the dashed lines) 
was divided by 8 -- the number of atoms per layer. 
The central layer magnetization is quite close to the 
theoretical\cite{mittendorfer} and experimental\cite{muexp} value $0.61~\mu_B$ for magnetic moment of bulk Ni.
The local  magnetization   
grows when moving from bulk-like internal layers to the surface layer and further
to the central atom of the nanowire. This reflects a common tendency in $d$-band metals 
(e.g., Ni, Co, and Fe) for the magnetic moment to be enhanced in conditions
of reduced coordination, as found in systems with reduced system dimensionality
where the $d$ bands are narrower. The magnetic moments of the atoms
at the tip surface and of those in interior layers ($\mu_{S1}=0.72~\mu_B$ and $\mu_{S2}=0.64~\mu_B$)
compare well with the corresponding values calculated for the Ni(001) surface\cite{mittendorfer} 
($0.76~\mu_B$ and $0.68~\mu_B$). Very reasonably, magnetization in the lead region
was found to be quite insensitive to the structural relaxation.
On the contrary, magnetic moments on the nanowire atoms reduce significantly during the 
relaxation as a consequence of a significant shortening of interatomic bonds. 
In particular, the magnetic moment on the central nanowire atom drops from $1.22~\mu_B$ 
to $1.07~\mu_B$. These values are very close to those ($1.19~\mu_B$ and $1.13~\mu_B$) 
for an infinite monatomic wire calculated at the corresponding interatomic distances.

\section{Electron Transmission through the Nanocontact}

Conductance is the main measurable physical quantity in nanocontacts. Owing to their small
size, usually well below the electron mean free path, resistive processes of diffusive
and of inelastic origin are generally negligible in nanocontacts, where even at room 
temperature the conductance is ballistic.
In the linear response regime (i.e., at very small applied voltages) the ballistic conductance 
of an open system consisting of the nanocontact connected to two semi-infinite leads is related
to the total transmission function at the Fermi energy 
by the Landauer-B\"uttiker formula: $G=(1/2)G_0 T(E_F)$, where $G_0=2e^2/h$ is the
conductance quantum. 
In our spin density functional picture electrons
of different spin move independently in their different self-consistent potentials.
The two spin channels are uncoupled and can be treated independently, so long as
the spin-flip processes are neglected, and the total transmission is the sum of 
contributions from the two spin channels: $T(E)=T_\uparrow(E)+T_\downarrow(E)$.
In our transmission calculation, we consider the system shown in Fig.~\ref{struc}
as the nanocontact scattering region, ideally attached to semi-infinite leads
on both sides. To obtain electron transmission at energy $E$ we first
construct the generalized Bloch states (consisting of propagating and evanescent 
states) which are solutions of the Kohn-Sham equation at energy $E$ 
of hypothetical infinite lattice-periodic leads.
These states are then used to construct the scattering states of the entire
lead-nanocontact-lead system and to calculate the transmission function. Full details
are given in Ref.[\onlinecite{ourprb}]. 

\begin{figure}
\includegraphics[width=8.5cm]{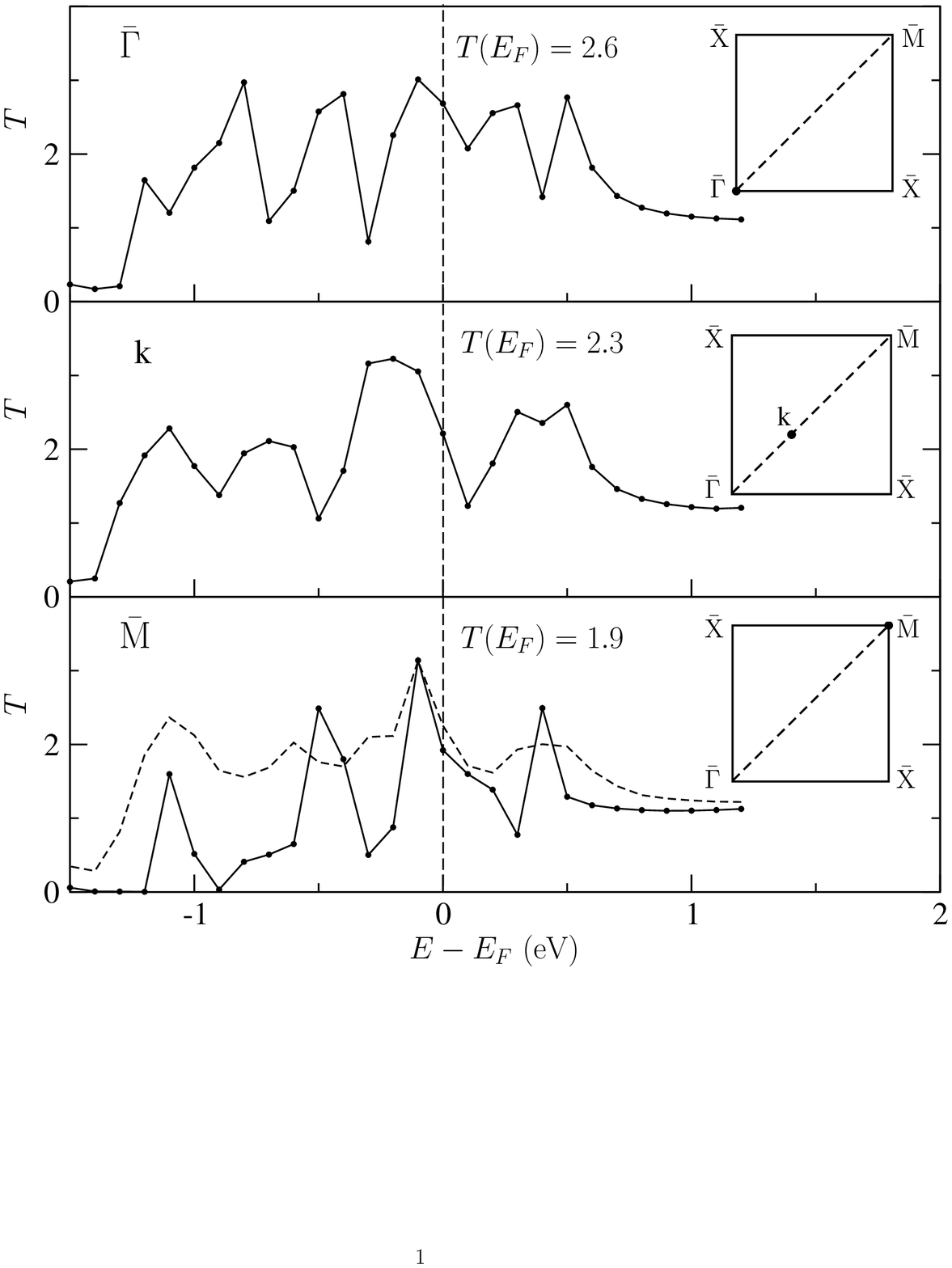}
\caption{\label{kdiff}
Transmission versus energy for the minority spin channel of the
(repeated) Ni nanowire in relaxed configuration, calculated at different $\kp$.
Selected $\kp$ values are shown on insets.
The numerical value of transmission at the Fermi energy -- representative of
ballistic conductance at zero voltage -- is also given for each curve.
The transmission function obtained by averaging over the full set of
10 special $\kp$ points, closest to the single nanowire conductance,
is plotted as the dashed curve on the lowest panel.
}
\end{figure}

Our supercell geometry encompasses an infinite number of nanowire replicas 
forming a square supercell lattice in the $xy$ plane. The transmission (per supercell)
of such infinite set of nanocontacts is given by an integral over the 2D BZ
of a transmission that depends on $\kp=(k_x, k_y)$:
\begin{equation}
T(E)={S\over (2\pi)^2}\int T(E,\kp)d^2\kp,
\end{equation}
where $S$ is the area of the supercell in the $xy$ plane.
What we are interested in, however, is the transmission
of a single isolated nanocontact; we 
should therefore try to attain the limit of an infinitely large supercell. 
One could approach this limit in many different ways, for example 
by calculating transmission at some $\kp$ (e.g., at the $\bar{\Gamma}$ point, 
$\kp=0$) or else by integrating the transmission over the 2D BZ as in Eq.~1.
For a sufficiently large supercell the 2D BZ will collapse to a point,
so that the transmission function will not depend on $\kp$ and 
the two approaches will eventually give the same limiting result.
We found, however, that upon increasing the supercell
$xy$ size the correct single nanocontact limiting value for 
transmission function is approached much faster in the second case.
In particular, for a manageable supercell size such as $(2\sqrt{2}\times2\sqrt{2})$ 
the $\kp$ point integration of the transmission function
is crucial since it turns out to depend much on the $\kp$,
as illustrated in Fig.~\ref{kdiff} where the minority spin channel transmission
of a three-atom Ni nanowire (in relaxed configuration) is presented for 
the different $\kp$ values shown in the insets. The transmission function varies 
significantly from one $\kp$ to another reflecting a different 
coupling between the nanowire and the lead states. In particular transmission at
the Fermi energy, which gives the conductance, changes quite noticeably with $\kp$.
The fully $\kp$-integrated transmission (obtained by using 10 special $\kp$ points)
is provided for comparison in the lowest panel (dashed curve). Here 
$\kp$-averaging makes the transmission curve smoother, eliminating
several fluctuations and retaining only smoother features attributable to the
isolated nanowire. We verified in fact that this $\kp$-integrated conductance 
depends much less on the supercell size than any of the selected $\kp$ points, 
and is thus the best approximation to that of the isolated nanowire.
The importance of proper sampling of the 2D BZ for ballistic conductance 
calculation was also pointed out recently by Thygesen {\it et al.}\cite{ksampling}
In conclusion we systematically integrated the transmission function over the 2D BZ
of a ($2\sqrt{2}\times2\sqrt{2}$) $xy$ supercell by averaging over 
10 special $\kp$ points in the irreducible sector of the 2D BZ.\cite{Al}

\section{Connection with Electronic Structure}

Before discussing the overall ballistic conductance results for the Ni nanowire
we wish to discuss the electronic structure of the supercell used to 
calculate the self-consistent potential.
As we shall show, at each $\kp$ there is a clear correlation between sharp features 
in the transmission as a function of energy and the electronic structure
of the supercell.
Let us consider in particular the 
dispersion of states with $k_z$. If the nanocontact were cut, all the 
states would be $z$-localized inside the lead block, forming a dense set of
states that are flat in $k_z$, totally devoid of dispersion. When 
the nanocontact is established, a majority of these states remain still
localized inside the lead, but a few extended states appear, characterized
by $k_z$ dispersion. These extended states are not necessarily easy to
identify, because as a function of $k_z$ they (anti-)cross the dense lead 
localized states. Yet, those extended states that cross $E_F$ are responsible
for the nanocontact ballistic conductance.   

\begin{figure}
\includegraphics[width=8.5cm]{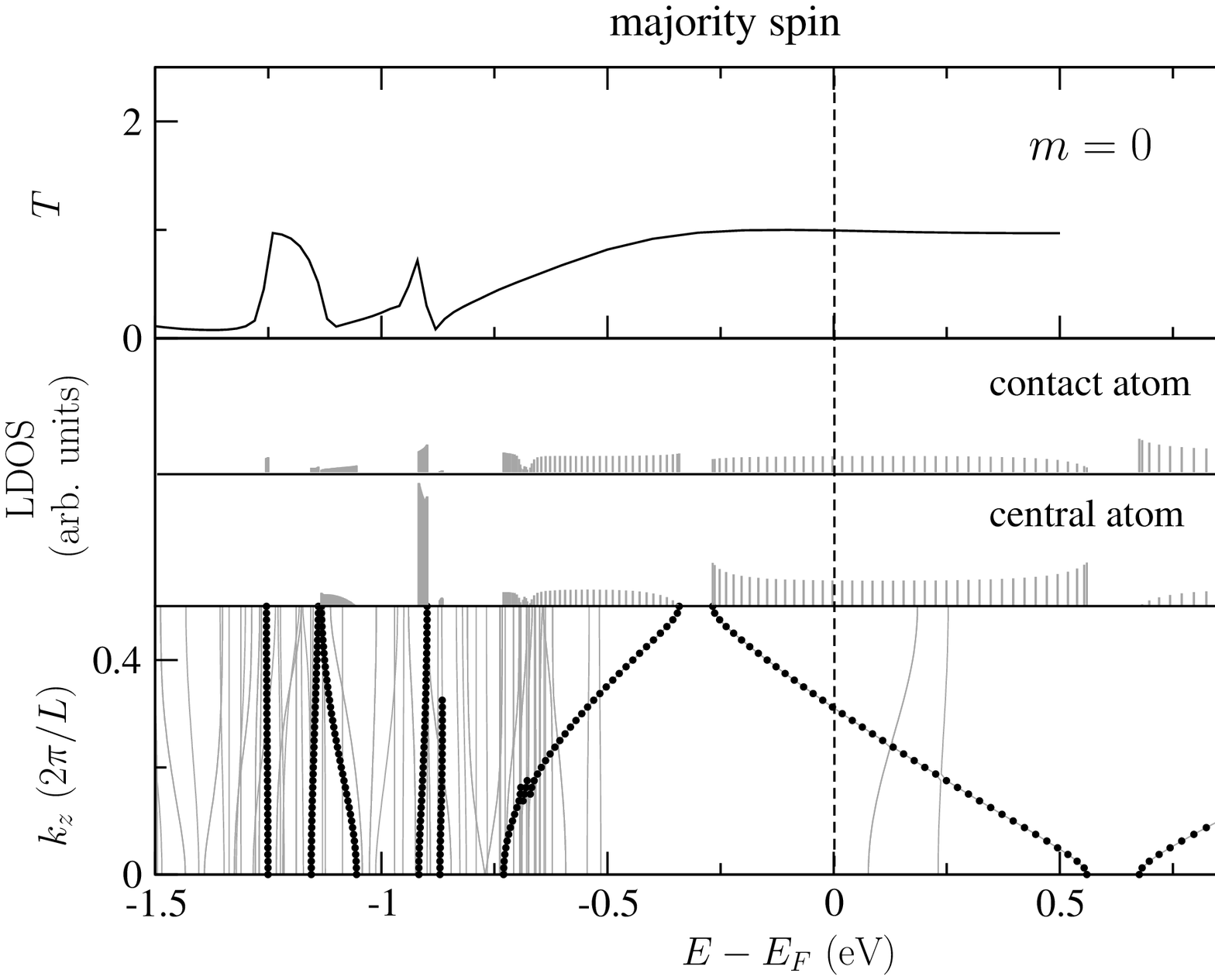}
\caption{\label{bandsup}
Supercell band structure for the majority spin of the relaxed three-atom Ni wire calculated along the
$z$ direction at the $\bar{\rm M}$ point of the 2D BZ.
Each electron state can be labeled by a certain $|m|$, where $m$ is the projection
of the angular momentum on the nanowire axis (see text).
The circles mark the states of $m=0$ symmetry having the charge on
the wire larger than some threshold value (see text) while
the states of $|m|=1$ and $|m|=2$ symmetry are not presented since they are irrelevant for
electron transport.
For each $m=0$ state, the charge on the contact and central atoms of the wire is given by vertical line
s
on the LDOS panel.
The electron transmission of the $m=0$ channel is also shown.}
\end{figure}

\begin{figure}
\includegraphics[width=8.5cm]{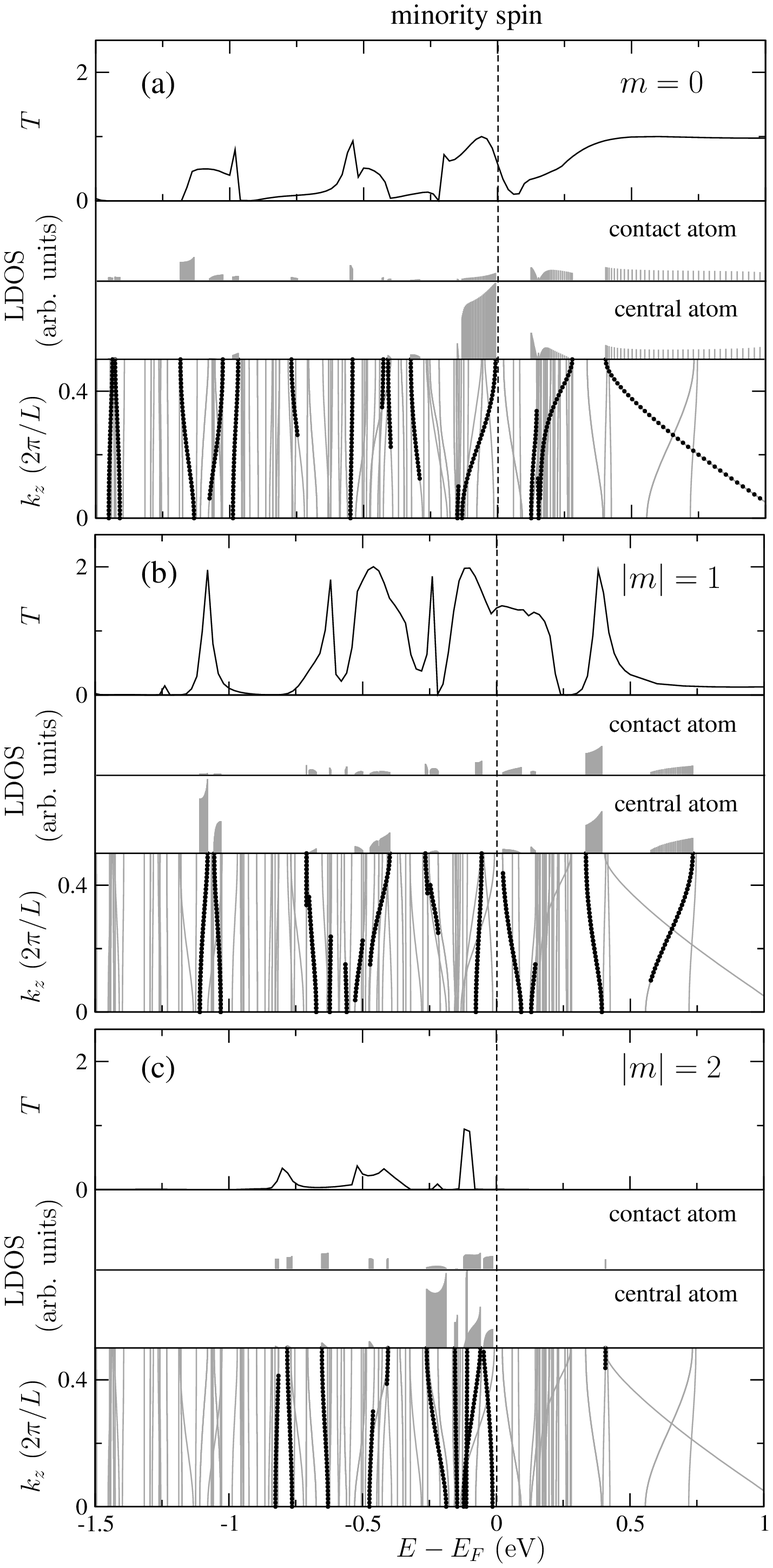}
\caption{\label{bandsdown}
The same as in Fig.~\ref{bandsup} but for the minority spin channel.
In addition to $m=0$ states (a), the corresponding data for the states of $|m|=1$ (b) and of $|m|=2$ (c
)
symmetry (which now contribute to the transmission around the Fermi energy) are also shown.}
\end{figure}

For illustration, we choose 
the $\bar{\rm M}$ point of the 2D BZ. In Figs.~\ref{bandsup} and \ref{bandsdown} we present  
electron bands of the three-atom Ni nanowire (relaxed configuration) in the $z$ direction
for majority and minority spins, respectively.
We classify the electron states in terms of the orbital angular momentum $m$ 
around the nanowire axis. 
Though in our geometry the $m$ 
is not strictly a good quantum number due to the presence of the leads,
all the electron states at the $\bar{\rm M}$ point (as well as at the $\bar{\Gamma}$ point)
can be separated into three groups which we call as $m=0$, $|m|=1$, and $|m|=2$.
We do this separation as follows.
At the $\bar{\rm M}$ point (and at the $\bar{\Gamma}$ point),
electron states are classified according to irreducible representations of the
group $C_{4v}$.
This group has four one-dimensional representations
$\Delta_1$, $\Delta_{1'}$, $\Delta_2$, and $\Delta_{2'}$
and one two-dimensional representation $\Delta_5$.
At nanowire atoms the $\{s-p_z-d_{3z^2-r^2}\}$, $\{p_x-d_{xz},p_y-d_{yz}\}$, $\{d_{x^2-y^2}\}$,
and $\{d_{xy}\}$ atomic orbitals transform according to $\Delta_1$, $\Delta_5$, $\Delta_2$, and $\Delta_{2'}$ representations, respectively.
Therefore, we can assign $m=0$ to states of $\Delta_1$,
$|m|=1$ to states of $\Delta_5$, and $|m|=2$ to states of $\Delta_2$ and $\Delta_{2'}$.
Note that 
there will be in principle smaller contributions from higher
atomic states such as $f$, $g$, but that is not very important.
Similarly, in a transmission calculation each transmission eigenchannel
(obtained by diagonalizing the Hermitian matrix ${\bf T}^+{\bf T}$, ${\bf T}$ being
the transmission matrix\cite{eigenchannels})
is clearly associated with a well defined $|m|$ so that the total transmission is naturally
decomposed onto contributions from $m=0$, $|m|=1$, and $|m|=2$. 

Let us consider majority spin first (Fig.~\ref{bandsup}). All the $\bar{\rm M}$ point electron states
as a function of $k_z$ are plotted by thin lines. Moreover, the states of $m=0$ symmetry  
are marked by circles when their squared wave-functions have 
a charge on the nanowire larger than 4\%. 
The amount of charge for these states on both the first contact atom and the nanowire
middle atom (central atom) is given by the height of vertical lines on the LDOS panels.
In Ni all spin up $d$ states are occupied so that the corresponding bands of $|m|=1$ and $|m|=2$
symmetry lie well below the Fermi energy. They are clearly
visible in Fig~\ref{bandsup} as the crowd of localized states amassed below energy
$-0.5$ eV.
These states should not participate to electron transport
and we do not display in Fig.~\ref{bandsup} the data for $|m|=1$ and $|m|=2$.
Around the Fermi energy the extended $m=0$ states through the wire have mainly $s-p_z$ character.
They give rise to a dispersive band as shown in Fig.~\ref{bandsup}. Accordingly, 
the transmission of the $m=0$ channel is a smooth function of energy and is close to 1,
because the $s$ band is broad, and with its high kinetic energy it is very poorly reflected
at the nanocontact. Below $-0.75$ eV, however, the $s-p_z$ states hybridize strongly 
with $d_{3z^2-r^2}$ states, making their effective dispersion much flatter, or
the effective mass along $z$ much bigger. 
As a result, in this energy region well below $E_F$ the $|m|$=0 transmission 
drops well below 1, developing a strong energy dependence.

The situation is considerably richer for minority spin (Fig.~\ref{bandsdown}). Here the $d$ states of 
the nanowire are partially occupied and contribute to transmission around the Fermi energy.
Therefore, in Fig.~\ref{bandsdown} we also present the panels for $|m|=1$ (b) and $|m|=2$ (c). 
We mark the states by circles if their charge on the nanowire is larger   
than 4\% (for $|m|=1$) and 16\% (for $|m|=2$).
For the $m=0$ channel, 
the dispersive $s-p_z$ band ends well above $E_F$ at 
an energy about $0.4$ eV.
The Fermi energy falls inside the energy region of strong $s-p_z-d_{3z^2-r^2}$ hybridization.
Here, for all $|m|$, the electron bands are quite narrow which leads
to significant energy variations of the corresponding transmission function.
As anticipated there is a clear correlation between sharp features in the transmission
as a function of energy and the presence of extended states.
For example, the maximum in the transmission for the 
$m=0$ channel just below the Fermi energy corresponds
to a slightly dispersive band with mainly $d_{3z^2-r^2}$ character.
Similarly, the maxima in the $|m|=1$ transmission at 
about $-1.1$, $-0.5$, $-0.1$, and $0.4$ eV agree with the extended 
bands seen at those energies. 

It should be emphasized, however, that 
there are some aspects in the transmission function which can not be
understood solely from the band structure analysis. For instance, the 
energy band of $m=0$ symmetry has a small gap at an energy
around $-0.3$ eV (majority spin) or $0.3$ eV (minority spin) where the 
transmission function remains continuous and does not drop to zero. This 
is because the transmission function is really calculated for an 
infinite lead-nanocontact-lead open system (and should be in principle compared
with the LDOS constructed via scattering states),
whereas the LDOS shown refers
to the nanocontact alone, closed onto itself via periodic boundary conditions in the $z$ direction.
The two will approach one another only when increasing conceptually 
the number of bulk planes included in the supercell to infinity. The 
energy bands will then continuously change so that the band gap discussed above 
(which is an artifact of the periodic boundary condition along the $z$ axis) will eventually 
disappear. We actually tested these conclusions using a simple tight-binding model.
Note that the only quantity obtained from supercell calculations   
and needed for evaluation of transmission function is the self-consistent potential
which turns out to converge much better (than the LDOS) with the length of the supercell.

One feature which is important to underline again with regard to the  $m=0$ 
minority spin channel is that its transmission at $E_F$ is well below 1, despite
a massive presence of $s$ states. This is clearly the result of a
geometry-induced hybridization of $s-p_z$ states 
-- by themselves highly conductive-- with the poorly conductive $d_{3z^2-r^2}$ states.
We expect this hybridization to be in fact more general than the assumed 3-atom nanowire
geometry, and the $s$-channel transmission reduction to carry over to all cases
where hybridization is allowed by symmetry. This observation will be useful later.

\section{Ballistic Conductance}

\begin{figure}
\includegraphics[width=8.5cm]{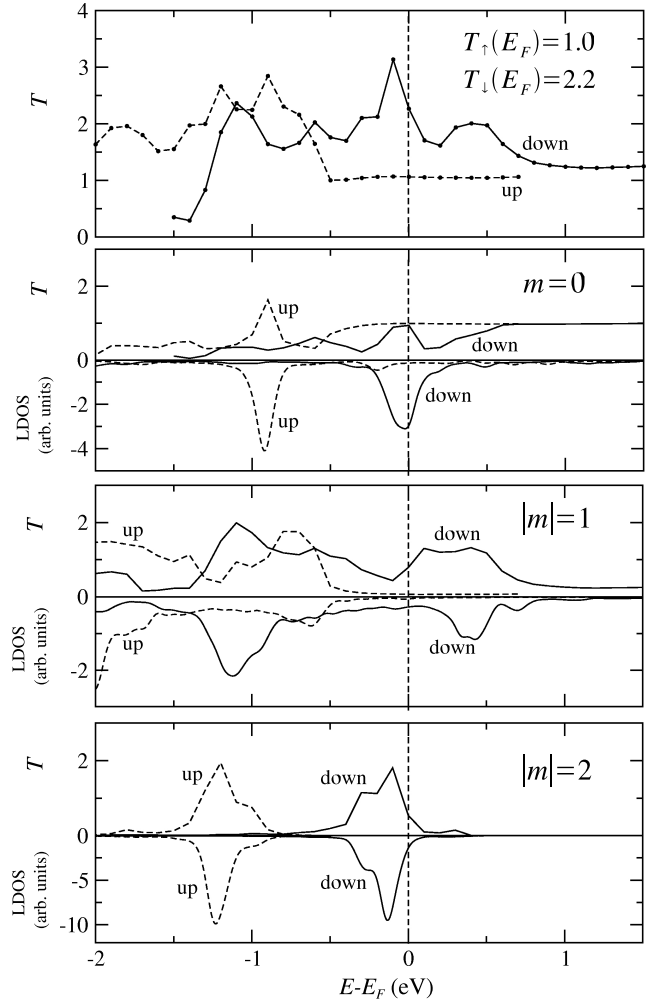}
\caption{\label{trelax}
Spin-dependent transmission function of the Ni nanowire in the fully
relaxed configuration (upper panel).
The lower panels display the symmetry decomposition of the transmission function and
of the LDOS at the mid-nanowire atom.
}
\end{figure}

We now present the ($\kp$-integrated) transmission function of the 3-atom suspended Ni 
nanowire in the relaxed configuration (Fig.~\ref{trelax}). 
On the upper panel we show the total transmission for the majority (up) and minority (down) spin channels.
To understand better the different features in the transmission curves we  
separate on the lower panels the total transmission onto contributions from different $|m|$
and also show the LDOS at the middle nanowire atom of the corresponding symmetry.
At arbitrary $\kp$ (unlike the $\bar{\rm M}$ or the $\bar{\Gamma}$ point), 
the transmission eigenchannels are not associated to a certain $|m|$ but have, in general, contributions from 
all the $|m|$. Therefore we decompose the total transmission as follows.
For each $\kp$ and for each transmission eigenchannel $j$ the 
charge inside a sphere of some radius $r_0$ (volume $V_0$) surrounding the middle nanowire 
atom is calculated as 
$n_j=\int_{V_0}|\psi_j({\bf r})|^2d{\bf r}$. It can be further decomposed onto contributions from
different $lm$ (in our calculations $l=0$, 1, and 2): 
$n_j=\sum_{lm} n_{j,lm}$, where $n_{j,lm}=\int_0^{r_0}|R_{j,lm}(r)|^2r^2dr$
and the projections  $R_{j,lm}(r)$ are obtained from 
$R_{j,lm}(r)=\int Y_{lm}^*(\Omega)\psi_j(r,\Omega)d\Omega$.
The transmission of each eigenchannel is then weighted by its charge of a certain symmetry.
For example, the contribution of the $j$-th channel to the transmission of $m=0$ symmetry
is given by $T_j(1/n_j)\sum_l n_{j,l0}$.
The contributions from different $\kp$ and from different eigenchannels 
are then properly added to produce the final value.

We first discuss the transmission function for the majority spin channel (spin up).
Here, the transmission around the Fermi energy is almost 1 and it is due to nearly 
free propagation of the $s-p_z$ state ($m=0$).
Since the spin up $d$ states of the wire are occupied, the $|m|=1$ and 
the $|m|=2$ channels contribute to the total transmission only at energies lower
than $E_F$ . Moreover, at $E_F$ the $|m|=1$ channel gives rise only to a small
tunneling current while the $|m|=2$ channel is completely closed.
For the minority spin channel (spin down), the transmission curves are 
qualitatively similar to majority spin channel but are shifted to higher 
energies due to the exchange gap, which in the Ni nanowire is $\sim$ 1.25 eV.
At the Fermi energy there are now down spin $d$ states available for transport.
The transmission of the $m=0$ channel at the Fermi energy is still close to 1 
but shows a strong energy dependence and corresponds to a mainly $d_{3z^2-r^2}$ 
LDOS peak at the central nanowire atom. 
The transmission function for the $|m|=1$ channel shows two broad maxima around
$-1.1$ eV and $0.4$ eV. They are associated with two double degenerate 
$d_{xz,yz}$ states of the three atom Ni wire. A third state is located 
at lower energy (about $-2.2$ eV) and is not shown in Fig.~\ref{trelax}. 
These three states will develop -- upon increasing the nanowire length 
-- into a twofold degenerate band with bandwidth about 3 eV.   
We find that the Fermi energy is located between two states of higher energy 
and the $|m|=1$ channel contributes 
here about 0.7 to the total transmission.
The remaining $|m|=2$ channel 
corresponds to a very narrow twofold degenerate $d_{xy,x^2-y^2}$
band of an infinite wire (the bandwidth is about 0.6 eV). This band, however, crosses the Fermi
energy. Therefore, right below the Fermi energy one has a noticeable 
contribution from the $|m|=2$ channel to the total transmission while its transmission
at the $E_F$ is about 0.5.
This channel is responsible for a transmission peak at the energy $-0.1$ eV.

\begin{figure}
\includegraphics[width=8.5cm]{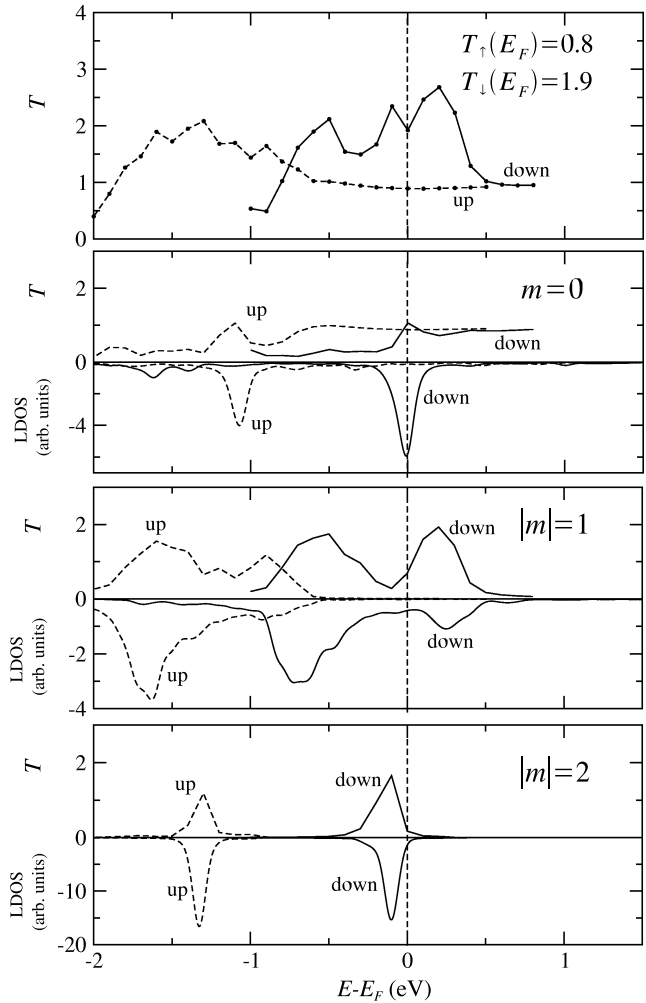}
\caption{\label{tunrelax}
The same as in Fig.~\ref{trelax} but for the Ni nanowire in unrelaxed configuration.
}
\end{figure}

We can further investigate how the structural relaxation affects electron 
transport. In Fig.~\ref{tunrelax} we present the transmission 
function of the Ni nanowire in the unrelaxed initial configuration. 
Similarly to the relaxed configuration (Fig.~\ref{trelax}), 
we also report the symmetry-decomposed transmission as well as the 
LDOS at the nanowire central atom.
One can see that the main conclusions drawn for the relaxed configuration 
remain unchanged.
The majority spin transmission 
at the Fermi energy is about 0.8 and is again dominated by the $s$-like channel.
Separating the minority spin transmission at the Fermi energy, one obtains values of 
1.0, 0.7, and 0.2 from $m=0$, $|m|=1$, and $|m|=2$ channels, respectively.
Due to larger wire-lead separation in the unrelaxed configuration (parameter $D$ in Table.~1),  
the nanowire states are less coupled to the leads. As a result, the peaks in transmission
curves get sharper (see, e.g., two peaks for the $|m|=1$ channel and the peak for the $|m|=2$ channel).
This leads to the smaller transmission at the Fermi energy for the $|m|=2$ channel -- 0.2 versus 0.5
in the relaxed configuration (see Fig.~\ref{trelax}).
For the $|m|=1$ channel, the longer interatomic bonds in the wire (parameter $d$ in Table.~1)
lead to the fact that the two maxima in the transmission are less separated in energy and move closer
to the Fermi level.
This compensates the previous effect so that the transmission
at the Fermi energy for the $|m|=1$ channel does not practically change compared to the
relaxed configuration.

We can now translate all results into
a calculated ballistic conductance. From the total transmission at the Fermi 
energy we obtain a conductance $G\approx(0.5+1.1)~G_0=1.6~G_0$ and 
$G\approx(0.4+0.95)~G_0=1.35~G_0$ for relaxed and unrelaxed nanowires, respectively.
The conductance for the relaxed configuration turns out to be larger with respect to the unrelaxed
case which is reasonable because of smaller lead-nanowire separation (better contact). 
The calculated conductances are in good agreement with break-junction measurements showing a 
quite broad first conductance peak  
centered at $1.6~G_0$\cite{yanson} or at $1.3~G_0$.\cite{untiedt} The breadth 
of the measured conductance histogram peak can be explained by observing 
that the detailed nanocontact conductance depends noticeably on the detailed
geometry of the lead-nanocontact structure, as was seen for example by comparison 
of the relaxed and unrelaxed three-atom geometries. 
Recently, Bagrets {\it et al.}\cite{bagrets} calculated the ballistic conductance of a three-atom Co nanowire suspended
between two Co(001) leads. They considered an unrelaxed nanowire (which 
corresponds to our unrelaxed configuration)
and reported values close to $0.5~G_0$ for both majority and minority spin conductances.
In comparison with our Ni results that finding is somewhat puzzling, as we obtain 
a much higher minority spin conductance 
due to a sizable contribution from the $d$ electrons. 
In this respect, our results are closer to recent conductance calculations performed
by Jacob {\it et al.}\cite{jacob} They studied nanocontacts formed by two Ni atoms joining the two Ni electrodes
and also found a non zero conductance from minority spin $d$ states  
at low nanocontact stretching.

We should also stress that conductances reported here  
are much smaller than the ideal one for an infinite tipless monatomic wire $G=3.5~G_0$, 
where in the absence  of any reflection the conductance is just equal 
to the number of bands at the Fermi energy which is 7. The reduction from this ideal
3.5 conductance to the real 1.6 is due to the fact that $d$ electrons are reflected
by about 75\%, and also to $s-d$ hybridization. On the other hand, this limited transmission
effect is not as drastic as the predicted effect of a magnetization reversal 
(or ``domain wall'') built into an infinite nanowire.\cite{ourprb} 
In the latter case all the $d$ channels are completely blocked at the 
Fermi energy so that conductance of the nanowire was only $G=1~G_0$ 
due to two (one per spin) $s$-like channels.

\section{Conclusions}
To summarize, we have presented DFT calculations of the structural and magnetic 
properties of the three-atom Ni nanowire
suspended between two semi-infinite Ni(001) leads. A scattering-based approach was applied to study ballistic
transport in both unrelaxed and relaxed configurations of the nanowire. 
Our calculations show that in both atomic configurations the majority spin conductance  
is attributed to one $s$-like channel and is, therefore, very close to $0.5~G_0$.
On the contrary, the minority spin conductance has also noticeable contribution from the $d$ states and
is higher -- $1.35~G_0$ and $1.6~G_0$ for unrelaxed and relaxed configurations,
respectively. We found that overall conductances are quite close to recent experimental break junction results. 
Moreover, the fact that conductance is sensitive to the details of underlying atomic structure could account 
for a quite broad first peak in conductance histograms observed in experiment.

\section{Acknowledgments}
This work has been supported by the Italian MIUR and
by INFM through ``Iniziativa trasversale calcolo parallelo''.


\begin{thebibliography}{99}
\bibitem{nota1}
The amount of contraction depends on the specific second layer atom
since not all atoms here are equivalent (unlike the surface plane atoms).
For example, the second layer atom beneath the nanowire end
atom moves noticeably outwards by about 0.05~\AA, or 2.8\% $d_{12}^0$ 
relative to the surface plane. Two other (mutually equivalent) atoms of the second 
layer located on the sides of the square $xy$ supercell move outwards by only about 1.6\% $d_{12}^0$ 
while the fourth second layer atom (located on the diagonal of the square $xy$ 
supercell) relaxes by only 0.8\% $d_{12}^0$.
\bibitem{nota2}
Additional atoms of the second layer lacking equivalent atoms in the ($2\times 2$) geometry 
were added at the same $z$ value as that of the fourth second layer atom described in 
Ref. \onlinecite{nota1}.
Four new atoms were also  added to form the ($2\sqrt{2}\times2\sqrt{2}$)
outermost surface layer in such a way that their distance to the added
second layer atoms was equal to the first interlayer distance calculated
for a relaxed Ni(001) surface, namely 1.7 \AA.
The atomic configuration so obtained was finally checked for residual
atomic forces less than 5 mRy/a.u.
\bibitem{Al}
In the specific example of a short Al nanowire suspended between two Al(001) leads,
we considered ($2\times2$), ($2\sqrt{2}\times2\sqrt{2}$), and ($4\times4$) supercell
geometries, and found that a reasonably converged transmission function is obtained
already with the $(2\sqrt{2}\times2\sqrt{2})$ $xy$ size, a $\kp$ integration
being performed using 10 special $\kp$ points. This $xy$ geometry and BZ sampling were
then adopted also for Ni nanowires.
\end{thebibliography}
\end{document}